\documentclass[final,times,5p,twocolumn]{elsarticle}

\usepackage{graphics}
\usepackage{epsfig}

\usepackage{amsthm}
\usepackage{graphicx,color}
\usepackage{amsmath,amssymb,amscd}
\usepackage[caption=false]{subfig}

\newcommand*{\hc}{\dagger}
\newcommand*{\pd}{\partial}
\newcommand*{\wt}{\widetilde}
\newcommand*{\ovl}{\overline}
\newcommand*{\ra}{\rightarrow}

\newcommand*{\Lra}{\Longrightarrow}

\newcommand*{\del}{\delta}

\newcommand*{\eps}{\epsilon}
\newcommand*{\gm}{\gamma}
\newcommand*{\lam}{\lambda}

\newcommand*{\half}{\frac{1}{2}}
\newcommand*{\lgn}{\mathcal{L}}
\newcommand*{\PL}{\hat{L}}
\newcommand*{\PR}{\hat{R}}

\journal{Physics Letters B}

\begin{document}
\title{Some consequences of the Majoron being the dark radiation}
\begin{frontmatter}
\author[label1,label2]{We-Fu Chang}
\ead{wfchang@phys.nthu.edu.tw}
\author[label3]{J. N. Ng}
\ead{misery@triumf.ca}
\author[label1,label2]{Jackson M. S. Wu}
\ead{jknw350@yahoo.com}
\address[label1]{Department of Physics, National Tsing Hua University, 
No. 101, Section 2, Kuang-Fu Road, Hsinchu, Taiwan 30013, R.O.C.}
\address[label2]{Physics Division, National Center for Theoretical Sciences, 
No. 101, Section 2, Kuang-Fu Road, Hsinchu, Taiwan 30013, R.O.C.}
\address[label3]{TRIUMF, 4004 Wesbrook Mall, Vancouver, BC, V6T 2A3, Canada}

\begin{abstract}
We discuss some phenomenological consequences in a scenario where a singlet Majoron plays the role of dark radiation. We study the interrelations between neutrino mass generation and the scalar potential arising from this identification. We find the extra scalar has to be light with a mass at or below the GeV level. The mixing of this scalar with the Standard Model Higgs impacts low energy phenomena such as the muonic hydrogen Lamb shift and muon anomalous magnetic moment. Demanding that the light scalar solves the puzzle in the muon magnetic moment requires the scalar to be lighter still with mass at or below the 10~MeV level. The 
cross-sections for the production of heavy neutrinos at LHC14 are also given.
\end{abstract}

\end{frontmatter}


\section{Introduction}

It is well known that correlations of temperature fluctuations in the Cosmic Microwave Background (CMB) depends
on the number of effective relativistic degrees of freedom, $N_{\text{eff}}$, which is usually given in terms of the
effective number of neutrinos species present in the era before recombination. The expected value of $N_{\text{eff}} = 3$ is consistent
with observations thus far. However, recent measurements of CMB from the Planck satellite~\cite{Planck} combined with that of the Hubble constant
from the Hubble Space Telescope (HST)~\cite{HST} resulted in a higher value of $N_{\text{eff}} = 3.83\pm 0.54$ at 95\%CL. If one further includes data from WMAP9~\cite{W9}, Atacama Cosmology Telescope (ACT)~\cite{ACT} and South Pole Telescope (SPT)~\cite{SPT} into the analysis, the extracted value becomes $N_{\text{eff}} = 3.62^{+0.50}_{-0.48}$ at 95\%CL. This hints at a dark radiation (DR) component beyond the expected three neutrino species at a confidence level of $2.4\sigma$. The origin and nature of such DR component is not known. One possibility, as pointed out recently by Weinberg~\cite{Weinberg:DR}, is that it can be naturally associated with a massless or nearly massless Goldstone boson arising the spontaneous breaking of a $U(1)$ global symmetry. A Goldstone boson will count as 4/7 of a neutrino, and this appears to agree with observation. However, in order for the temperature of the Goldstone bosons to match with that of the neutrinos, they must remain in thermal equilibrium with ordinary matter 
until muon annihilation. If Goldstone bosons decouple much earlier, they will contribute less than 4/7 to $N_{\text{eff}}$ as they will not be reheated but the neutrinos do. Decoupling in the muon annihilation era yields a contribution $\del N_{\text{eff}} = 0.39$. Weinberg further proposed that the $U(1)$ global symmetry be a new one associated with a dark sector with its own matter content. Decoupling in the muon annihilation era yields a contribution $\del N_{\text{eff}} = 0.39$. Weinberg further proposed that the $U(1)$ global symmetry be a new one associated with a dark sector with its own matter content.

In this letter we examine the possibility of taking the global $U(1)$ symmetry to be the lepton number. The spontaneous breaking
of this $U(1)_L$ by singlet Higgs will give rise to a Majoron~\cite{Peccei:SingletMajoron}, which we associate with the Goldstone boson
that acts as DR. Since the singlet that breaks the $U(1)_L$ will mix with the Standard Model (SM) Higgs boson,
this allows us to connect Higgs physics and DR to neutrino physics. In particular, we are able to link constraints on the parameters of the scalar sector to that in the seesaw mechanism responsible for neutrino mass generation, and to study their interrelations. We illustrate this in Type-I seesaw~\cite{seesawI} and inverse seesaw~\cite{inseesaw} scenarios in this letter.

The organisation of this letter is as follows. In Sec.~\ref{sec:fmw}, we describe in detail the framework we use to study the interrelation between the Majorons, the neutrinos, and the scalars. In Sec.~\ref{sec:LEC}, we discuss some consequences on the scalar and neutrino parameters from measurements of the muon magnetic moment, Lamb shift of the muonic hydrogen, and decay rate of $\mu \ra e\gm$. In Sec.~\ref{sec:DY}, we evaluate the range of values of heavy neutrino masses and mixings that can be probed at the LHC. We end with a summary in Sec.~\ref{sec:summ}.

\section{\label{sec:fmw} The framework}
The simplest Majoron model extends the Standard Model (SM) by three generations of singlet righthanded (RH) neutrinos, $N_R^i$, and a singlet complex scalar, $S$~\cite{Peccei:SingletMajoron}. The relevant Yukawa interactions read
\begin{equation}\label{eq:yukss}
\lgn \supset -y_1\ovl{L_L}\wt{H}N_R - y_2\ovl{N_R^{\,c}} N_R S + h.c.
\end{equation}
where $L= (n_L,e_L)^T$ is the lefthanded (LH) SM lepton doublet, and $\wt{H} = i\sigma_2 H^\ast$ with $H$ the SM Higgs; the generation indices are suppressed for clarity. Note that there is an accidental global $U(1)$ symmetry associated with the conservation of lepton number ($L$) before electroweak symmetry breaking (EWSB) if $S$ is defined to have $L = -2$. After EWSB, we can write $H = (0, (v + h)/\sqrt{2})^T$ in the unitary gauge, where $v = 246.221$~GeV, and
\begin{equation}
S(x) = \frac{1}{\sqrt{2}}\left(v_S + s(x)\right)e^{2i\alpha(x)} \,.
\end{equation}
The Yukawa interactions~\eqref{eq:yukss} then give rise to neutrino masses, which take the form
\begin{equation}
\lgn \supset -
\begin{pmatrix}
\ovl{n_L} & \ovl{N_R^{\,c}}
\end{pmatrix}
\begin{pmatrix}
0   & m_D \\
m_D & M
\end{pmatrix}
\begin{pmatrix}
n_L^c \\
N_R
\end{pmatrix}
+ h.c. \,,
\end{equation}
where $m_D = 2^{-3/2} y_1 v$, $M = y_2 v_S/\sqrt{2}$, and we have redefined the lepton fields, $\psi_l \ra e^{-i\alpha}\psi_l$,
to remove the $e^{2i\alpha}$ phase factor from the Majorana mass terms. Note that this induces the interactions $\pd_\mu\alpha\,\bar{\psi}_l\gm^\mu\psi_l$ from the lepton kinetic terms $\bar{\psi}_l\gm^\mu\pd_\mu\psi_l$.

If $\eps = m_D/M \ll 1$, the standard Type-I seesaw is operative. It is well known in this case that the light active neutrinos have masses $m_\nu \simeq \eps m_D$. To have $m_\nu \lesssim 0.1$~eV, we require
\begin{equation}\label{eq:sscond}
y_1 = 2^{5/4}\left(\frac{m_\nu y_2 v_S}{v}\right)^{1/2} \lesssim 3.05 \times 10^{-6}\left(\frac{y_2 v_S}{\mathrm{TeV}}\right)^{1/2} \,.
\end{equation}
As a benchmark, take $v_S = 1$~TeV and $y_2 = 1$. Then acceptable light neutrino masses can be obtained with $y_1$ the size of the electron Yukawa couplings, $y_e$. We shall refer to couplings with sizes smaller than $y_e$ as excessively fine tuned.

Type-I seesaw is not the only way neutrino masses can be generated, however. A phenomenologically more interesting case is the inverse seesaw. The inverse seesaw can be implemented by adding -- in addition to the three RH singlet neutrinos -- three more LH singlet neutrinos. The relevant Yukawa interactions given in Eq.~\eqref{eq:yukss} are now augmented to
\begin{equation}\label{eq:yukiss}
\lgn \supset -y_1\ovl{L_L}\wt{H}N_R' - y_2^R\ovl{N_R^{\prime\,c}}N_R' S - y_2^L\ovl{N_L^{\prime\,c}}\,N'_L S - M_D\ovl{N'_L} N'_R + h.c.
\end{equation}
where $N_{L,R}'$ are the LH and RH singlet neutrinos, and $M_D$ is a Dirac mass parameter. Note that a Yukawa coupling of the form $\ovl{L_L} H N_L^{\prime\,c}$ is forbidden by the global $U(1)$ lepton number. As above, the scalar phase can be removed by the appropriate lepton field redefinitions.

Mass terms arises after EWSB:
\begin{equation}
\label{eq:invseesaw}
\lgn \supset
\begin{pmatrix}
\ovl{n_L} & \ovl{N'_L} & \ovl{N_R^{\prime\,c}}
\end{pmatrix}
\begin{pmatrix}
0   & 0     & m_D \\
0   & \mu_L & M_D \\
m_D & M_D   & \mu_R
\end{pmatrix}
\begin{pmatrix}
n_L^c \\
N_L^{\prime\,c} \\
N'_R
\end{pmatrix}
+ h.c.
\end{equation}
where $m_D = 2^{-3/2}y_1 v$ and $\mu_{L,R} = y_2^{L,R}v_s/\sqrt{2}$. For $m_D,\,\mu_{L,R} \ll M_D$, it is useful to first go to a basis where large quantities are on the diagonal:
\begin{equation}
\begin{pmatrix}
N'_L \\
N_R^{\prime\,c}
\end{pmatrix}
= \frac{1}{\sqrt{2}}
\begin{pmatrix}
 1 & 1 \\
-1 & 1
\end{pmatrix}
\begin{pmatrix}
N_L \\
N_R^c
\end{pmatrix} \,.
\end{equation}
The mass matrix in the basis $\left\{n_L,\,N_L,\,N_R^c\right\}$ is then
\begin{equation}
\begin{pmatrix}
 0            & -m_D/\sqrt{2} & m_D/\sqrt{2} \\
-m_D/\sqrt{2} & -M_D + \mu_+  & \mu_- \\
 m_D/\sqrt{2} & \mu_-         & M_D + \mu_+
\end{pmatrix} \,,
\end{equation}
where $\mu_\pm = (\mu_L\pm\mu_R)/2$. To leading order in $\eps_D = m_D/M_D$ and $\eps_\pm = \mu_\pm/M_D$, the mass eigenstates are given by
\begin{align}
\nu_L &= n_L - \frac{\eps_D}{\sqrt{2}}N_L - \frac{\eps_D}{\sqrt{2}}N_R^c \,, \\
\eta_{1L} &= N_L + \frac{\eps_D}{\sqrt{2}}n_L - \frac{\eps_-}{2}N_R^c \,, \\
\eta_{2R} &= N_R + \frac{\eps_D}{\sqrt{2}}n_L^c + \frac{\eps_-}{2}N_L^c \,,
\end{align}
with mass eigenvalues $\eps_D^2\,\mu_+$, $M_D - \mu_+$, and $M_D + \mu_+$ respectively (after appropriate phase rotations).

The interactions between the Majoron and neutrinos arise from the neutrino kinetics terms. To leading order in $\eps_{D,\pm}$, they read
\begin{align}
\frac{\pd_\mu\chi}{2v_s}\Big[
&\bar{\nu}\gm^\mu\PL\nu + \bar{\eta}_1\gm^\mu\PL\eta_1 + \bar{\eta}_2\gm^\mu\PR\eta_2 \notag \\
-&\frac{\eps_D}{\sqrt{2}}\left(\bar{\eta}_2\gm^\mu\gm^5\nu +\bar{\nu}\gm^\mu\gm^5\eta_2\right)
-\frac{\eps_-}{2}\left(\bar{\eta}_2\gm^\mu\gm^5\eta_1 + \bar{\eta}_1\gm^\mu\gm^5\eta_2\right)
\Big] \,,
\end{align}
where $\PL$ and $\PR$ are the LH and RH chiral projectors. Note the absence of the $\chi$-$\nu$-$\eta_1$ coupling. Similarly, neutrino weak interactions in the mass eigenbasis read
\begin{align}
\lgn_{NC} &= \frac{g_W Z_\mu}{2\cos\theta_W}\Big\{
\bar{\nu}\gm^\mu\PL\nu + \frac{\eps_D^2}{2}(\bar{\eta}_1 + \bar{\eta}_2)\gamma^\mu\PL(\eta_1 + \eta_2) \notag \\
&\qquad\quad\quad + \frac{\eps_D}{\sqrt{2}}\left[\bar{\nu}\gm^\mu\PL(\eta_1+\eta_2) + (\bar{\eta}_1 + \bar{\eta}_2)\gm^\mu\PL\nu\right]
\Big\} \,, \\
\lgn_{CC} &= \frac{g_2}{\sqrt{2}}W^+_\mu
\left[\bar{\nu}\gamma^\mu\PL e + \frac{\eps_D}{\sqrt{2}}(\bar{\eta}_1 + \bar{\eta}_2)\gamma^\mu\PL e\right] + h.c.
\end{align}

Consider now the scalar sector. The most general renormalizable Lagrangian for it is given by
\begin{align}\label{eq:scalarV}
\lgn_{scalar} &= -(D_\mu H)^\hc(D^\mu H) - \pd_\mu S^\hc\pd^\mu S - V(H,S) \,, \\
V(H,S) &= -\mu^2 H^\hc H + \lam(H^\hc H)^2\notag \\
&\quad -\mu_S^2 S^\hc S  + \lam_S(S^\hc S)^2 + \lam_{HS}(H^\dag H)(S^\dag S) \,.
\end{align}
After EWSB, the kinetic term for $S$ takes the form
\begin{align}
\pd_\mu S^\hc\pd^\mu S &= \half\pd_\mu s\,\pd^\mu s + 2(v_S + s)^2\pd_\mu\alpha\pd^\mu\alpha \notag \\
&= \half\pd_\mu s\,\pd^\mu s + \half\pd_\mu\chi\pd^\mu\chi 
+ \left(\frac{s}{v_S} + \frac{s^2}{2v_S^2}\right)\pd_\mu\chi\pd^\mu\chi \,,
\end{align}
and we identify the canonically normalized Goldstone boson $\chi \equiv 2v_s\alpha$ as the Majoron.

The mixing between the Higgs doublet and the complex singlet was already analysed in Ref.~\cite{CNW07}. The classical minimum is given by
\begin{equation}
v^2 = \frac{4\lam_S\mu^2 - 2\lam_{HS}\mu_S^2}{4\lam\lam_S -\lam_{HS}^2} \,, \quad
v_S^2 = \frac{4\lam\mu_S^2 - 2\lam_{HS}\mu^2}{4\lam\lam_S -\lam_{HS}^2} \,.
\end{equation}
Using this, the scalar mass-squared matrix reads in the $(h,s)$ basis
\begin{equation}
\begin{pmatrix}
2\lam v^2      & \lam_{HS}v v_S \\
\lam_{HS}v v_S & 2\lam_S v_S^2
\end{pmatrix}
\end{equation}
which has eigenvalues
\begin{equation}\label{eq:mevals}
m_{1,2}^2 = \lam v^2 + \lam_S v_S^2 \mp \sqrt{(\lam_S v_S^2 - \lam v^2)^2 + \lam_{HS}^2 v^2 v_S^2} \,.
\end{equation}
The physical mass eigenstates are then
\begin{equation}
\begin{pmatrix}
h_1 \\
h_2
\end{pmatrix}
=
\begin{pmatrix}
\cos\theta & -\sin\theta \\
\sin\theta & \cos\theta
\end{pmatrix}
\begin{pmatrix}
h \\
s
\end{pmatrix} \,,
\end{equation}
with mixing angle
\begin{equation}\label{eq:thmix}
\tan 2\theta = \frac{\lam_{HS}v v_S}{\lam_S v_S^2 - \lam v^2} \,.
\end{equation}
We shall identify $h_1 \equiv h_{SM}$ as the SM Higgs, which was recently discovered at the LHC to have a mass of 125~GeV. Note that for small mixing (which shall be the case below), $m_{h_{SM}}^2 \approx 2\lam v^2$ and $m_2^2 \approx 2\lam_S v_S^2$.

From Eqs.~\eqref{eq:mevals} and~\eqref{eq:thmix}, the scalar quartic couplings can be written in terms of the mass eigenvalues and the mixing angle
\begin{align}
\label{eq:l2m}
\lam &= \frac{1}{4v^2}\left[m_1^2 + m_2^2 - (m_2^2 - m_1^2)c_{2\theta}\right] \,, \\
\label{eq:ls2m}
\lam_S &= \frac{1}{4v_S^2}\left[m_1^2 + m_2^2 + (m_2^2 - m_1^2)c_{2\theta}\right] \,, \\
\label{eq:lsh2m}
\lam_{HS} &= \frac{m_2^2 - m_1^2}{2v v_S}s_{2\theta} \,,
\end{align}
and we define the short hand $c_x\equiv\cos{x}$ etc. Classical stability of the vacuum demands that
\begin{equation}
\lam ,\, \lam_S > 0 \,, \qquad 4\lam\lam_S - \lam_{HS} = \frac{m_1^2 m_2^2}{v^2 v_S^2} > 0 \,,
\end{equation}
and we see from above that these conditions are automatically satisfied for $m_{1,2}$ real and positive.

Because of the scalar mixing, the SM Higgs can decay into a pair of Majorons. The partial width is given by
\begin{equation}\label{eq:h2inv}
\Gamma_{h_{SM}\ra\chi\chi} = \frac{s_\theta^2 m_{h_{SM}}^3}{32\pi v_S^2} \,.
\end{equation}
From the LHC, the Higgs invisible decay branching ratio is about 19\%~\cite{GKMRS13}. With the Higgs width at about 4.1~MeV~\cite{tome}, this means that
\begin{equation}\label{eq:vslb}
\Gamma_{h_{SM}\ra\chi\chi} \lesssim 0.8\,\mathrm{MeV} \Lra \frac{v_S}{|s_\theta|} \gtrsim 4.93\,\mathrm{TeV} \,.
\end{equation}
Currently, the LHC data on Higgs gauge boson couplings is consistent with SM expectations, which suggests small mixings with possible extended scalar sectors beyond the SM. As a benchmark, we take $s_\theta^2 < 0.1$, and we get $v_S \gtrsim 1.5$~TeV. Note that this lower bound on $v_S$ is relaxed if the bound on the mixing angle, $\theta$, becomes more stringent.

The scalar mixing at tree-level also give rise to the following effective interaction between the Majoron and the SM fermions:
\begin{equation}
\lgn_{ff\chi\chi}
= -\frac{\lam_{HS}m_f}{m_{h_{SM}}^2 m_{h_2}^2}\bar{f}{f}\pd_\mu\chi\pd^\mu\chi \,.
\end{equation}
As pointed out in Ref.~\cite{Weinberg:DR}, if the Majoron is to play the role of dark radiation that give rise to the fractional value of $N_\mathrm{eff}$ measured, it should stay in thermal equilibrium until roughly the time when muon annihilation happens. Then this requires the collision rate of Majorons with muons to be roughly the Hubble expansion rate:
\begin{equation}\label{eq:gb2DR}
\frac{\lam_{HS}^2 m_\mu^7 m_{Pl}}{m_{h_{SM}}^4 m_{h_2}^4} \approx 1 \Lra m_{h_2} \approx 9.3\sqrt{|\lam_{HS}|}\,\mathrm{GeV} \,,
\end{equation}
where we take $m_{h_{SM}} = 125$~GeV. With the help of Eqs.~\eqref{eq:lsh2m} and~\eqref{eq:h2inv}, we get from Eq.~\eqref{eq:gb2DR}
\begin{equation}
m_{h_2}^2 \approx \sqrt{X}\frac{1 - Y}{1 - Y^2} \,, \;\; Y = \frac{\sqrt{X}}{m_{h_{SM}}^2} \,, \;\;
X = \frac{32\pi\Gamma_{h_{SM}\ra\chi\chi}}{m_{h_{SM}}^3}\frac{c_\theta^2}{v^2}m_\mu^7 m_{Pl} \,.
\end{equation}
Then with $\Gamma_{h_{SM}\ra\chi\chi} \lesssim 0.8$~MeV, we obtain $m_{h_2} \lesssim 1.05$~GeV from $c_\theta < 1$, and from Eq.~\eqref{eq:ls2m} and the benchmark $c_\theta^2 > 0.9$:
\begin{equation}\label{eq:DRC}
\lam_S \lesssim 7.82 \times 10^{-4}\left(\frac{\mathrm{TeV}}{v_S}\right)^{2} \,.
\end{equation}
We see that in order to have no excessive fine tuning in the couplings, the lepton number breaking scale (as given by $v_S$) should be in the range of 1 to 30~TeV. On the other hand, it is well known that Type-I seesaw scenarios generally prefer a much higher scale, so $v_S$ should be very large, which then requires an excessive tuning of $\lam_S$. There is thus a tension between the very high scale Type-I seesaw and the identification of Majoron as DR in such scenarios.

Such tension, however, can be circumvented in the inverse seesaw scenario. As is seen in Eq.~\eqref{eq:invseesaw}, there are two scales that control the size of active neutrino masses, viz. $M_D$ and $v_S$. Without pushing $\lam_S$ to the nonperturbative region, we can take $v_S$ to be as low as $\mathcal{O}(10)$~GeV and still easily have $m_\nu \lesssim 0.1$~eV for the active neutrinos. For example, take $y_2 \sim 10^{-5}$, then $M_D$ can be as low as a few hundred GeV as long as $\eps_D \sim 10^{-3}$. This is an interesting region for LHC to look for heavy neutrinos that mix with the active ones, which we explore below in Sec.~\ref{sec:DY}. We note that low values of $v_S$ imply small mixings with the SM Higgs.

\section{\label{sec:LEC} Consequences from low energy physics}
\subsection{Muon magnetic moment and Lamb shift}
\begin{figure}[htbp]
\centering
\includegraphics[width=3.4in]{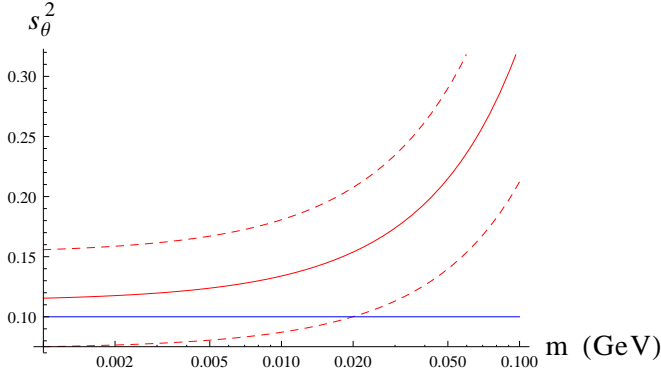}
\caption{Contribution to the muon magnetic moment due to the scalar $h_2$. The red solid curve gives the current central value of $\delta a_\mu$, while the red
dashed curves above and below it one sigma deviations above and below respectively. The horizontal blue line marks the benchmark LHC upper bound on the mixing angle, $s_\theta^2 = 0.1$}
\label{fig:phidamu}
\end{figure}

Due to scalar mixing, the light extra scalar $h_2$ has coupling $c_S^\mu = -2^{3/4}G_F^{1/2}m_\mu\,s_\theta$ to the muon arising from the Higgs Yukawa interactions with the fermions and is directly proportional to $s_\theta$. Its contribution to the muon magnetic moment is given by~\cite{amu}
\begin{equation}
\delta a_\mu = \frac{(c_S^\mu)^2}{8\pi^2}\int_0^1\!dz\,\frac{z^2(2-z)}{z^2 + r(1-z)} = \frac{(c_S^\mu)^2}{8\pi^2}H_S(r) \,,
\end{equation}
where $r = m_{h_2}^2/m_\mu^2$, and
\begin{align}
H_S(r) &= \frac{3}{2} - r + \frac{r(r-3)}{2}\log{r} \notag \\
&\quad - (r - 1)\sqrt{r(r-4)}\log\frac{\sqrt{r} + \sqrt{r - 4}}{2} \,.
\end{align}
Currently, the discrepancy between the experimental and theory value of $a_\mu = (g - 2)_\mu/2$ is~\cite{PDGmu}
\begin{equation}
\delta a_\mu = (249 \pm 87) \times 10^{-11} \,.
\end{equation}
We show in Fig.~\ref{fig:phidamu} the parameter space that this is accounted for by the $h_2$ contribution.
We see that only when $m_{h_2} < 0.02$~GeV is the benchmark bound on the mixing angle satisfied. We see also that the benchmark allowed parameter space has $\delta a_\mu$ lower than its current central value.

\begin{figure}[htbp]
\centering
\includegraphics[width=3.4in]{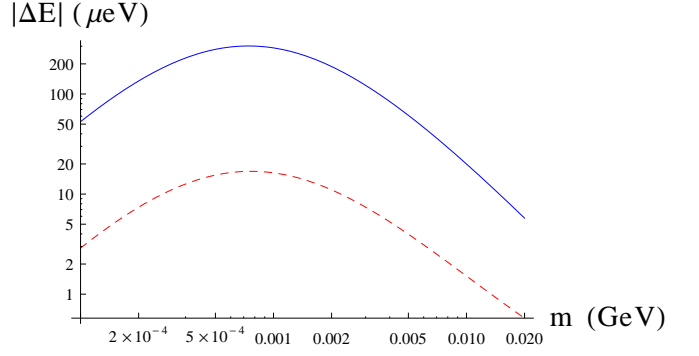}
\caption{Magnitude of the Lamb shift as a function of the scalar mass. The blue solid curve gives the shift at constant $s_\theta^2 = 0.1$, the red dashed curve that when the shift in the muon magnetic moment is kept constant at $\delta a_\mu = 162 \times 10^{-11}$, one sigma below the central value.}
\label{fig:dELS}
\end{figure}

Given the benchmark allowed parameter space, we can work out how much the additional scalar, $h_2$, contributes to the Lamb shift in muonic hydrogen. The energy difference from the $2P$-$2S$ splitting in hydrogen is given by~\cite{CR12,TSY10,Pachucki96}
\begin{equation}
\Delta E = -\frac{c_S^\mu c_S^p}{4\pi}\frac{m_{h_2}^2(m_r\alpha)^3}{2(m_{h_2} + m_r\alpha)^4} \,,
\end{equation}
where $m_r = m_\mu m_p/(m_\mu + m_p)$ is the reduced mass, and $c_S^{p} = -2^{3/4}G_F^{1/2}m_{p}\,\zeta\,s_\theta$ is the effective Yukawa couplings of $h_2$ to the proton, with $\zeta = 0.3 \sim 0.5$~\cite{HHG}. Fig.~\ref{fig:dELS} shows the magnitude of this energy shift in the parameter space allowed by both
$\delta a_\mu$ and our LHC benchmark.
We see that the maximum Lamb shift coming from $h_2$ alone is about 210~$\mu$eV, and this requires a very light $h_2$ with mass about 1~MeV. Although this is a significant portion of the 310~$\mu$eV needed to reconcile the current 7$\sigma$ discrepancy between the Committee on Data for Science and Technology (CODATA) value on the proton charge radius~\cite{CODATA2010} -- which is determined purely from electron scatterings -- and that measured from muonic hydrogen Lamb shift~\cite{muonLS}, it is not enough on its own. To solve the proton radius puzzle, additional ingredients besides $h_2$ is necessary.

\subsection{The radiative decay $\mu\ra e\,\gamma$}
The radiative $\mu \ra e\,\gamma$ decay here is mediated by both $W$ and heavy Majorana neutrino exchanges at one-loop. We can place limits on the neutrino mixings from the branching ratio of this decay. The gauge-invariant effective operator for $\mu\ra e\,\gamma$ has the form
$\frac{1}{M^2}\bar{L}{\sigma^{\mu\nu}}e H F_{\mu\nu}$, with the heavy neutrino mass, $M$, the controlling scale here. For simplicity, we assume that the
$\eta l^\pm W^\mp$ couplings are flavor universal given by $\eps\,g_2 /\sqrt{2}$. We shall use $\eta$ and $\eps$ here and below to denote generically the heavy Majorana neutrinos and their mixings with the light neutrino respectively (e.g. $\eps = m_D/M$ for Type-I seesaw, $\eps = \eps_D/\sqrt{2}$ for inverse seesaw). The effective Lagrangian then reads
\begin{equation}
\frac{e\,g_2^2\,\eps^2}{16\pi^2} \frac{m_\mu}{M^2} \left(\bar{e}\sigma^{\mu\nu}\PR\mu F_{\mu\nu}\right) \,.
\end{equation}
Following Ref.~\cite{WFJ_LFV}, the corresponding dipole coefficient is estimated as
\begin{equation}
A_L\simeq\frac{e\,g_2^2\,\eps^2}{32\sqrt{2}\pi^2}\frac{1}{G_F M^2} = \frac{e\,\epsilon^2}{8\pi^2} \frac{M_W^2}{M^2} \,,
\end{equation}
and thus the branching ratio of $\mu\ra e \gamma$
\begin{equation}
Br(\mu\ra e\,\gamma)=384\pi^2 |A_L|^2 \simeq 24\left( \frac{\alpha}{\pi}\right) \epsilon^4 \frac{M_W^4}{M^4} \leq 10^{-12} \,,
\end{equation}
which implies
\begin{equation}
\epsilon \lesssim 0.0142 \left(\frac{M}{\mathrm{TeV}}\right) \,.
\end{equation}
This constraint can be easily accommodated in both neutrino mass models.

\section{\label{sec:DY} Drell-Yan production of heavy neutrinos at the LHC}
At the parton level, the Drell-Yan (DY) production of the heavy Majorana neutrino, $\eta$, at the LHC proceeds predominantly through two processes:
$ q(p_1) + \bar{q}(p_2) \ra Z^* \ra \eta +\bar{\nu}$ and $u(p_1) + \bar{d}(p_2) \ra W^* \ra \eta + \bar{l}$.
The parton level cross-sections read
\begin{align}
\hat{\sigma}_Z(\hat{s}) & =
\frac{g_L^2+g_R^2}{384\pi}\left(\frac{g_2}{c_W}\right)^4|\eps|^2
\frac{1}{\hat{s}}\left(\frac{\hat{s} -m_\eta^2}{\hat{s}-M_Z^2}\right)^2
\left(1+\frac{m_\eta^2}{2\hat{s}}\right) \,, \\
\hat{\sigma}_W(\hat{s}) &=
\frac{ g_2^4}{384\pi}|\eps|^2 
\frac{1}{\hat{s}}\left(\frac{\hat{s} -m_\eta^2}{\hat{s}-M_W^2}\right)^2\left(1+\frac{m_\eta^2}{2\hat{s}}\right) \,,
\end{align}
where $g_{L,R}=T^3_{L,R}-Q_{L,R}\sin^2\theta_W$, $\eps$ is the heavy-light neutrino mixing, and $\hat{s}=(p_1+p_2)^2=x_1 x_2 s$ with $s$ the center-of-mass (CM) energy, and $x_{1,2}$ the parton momentum fractions.

\begin{figure}[htbp]
\centering
\includegraphics[width=3.45in]{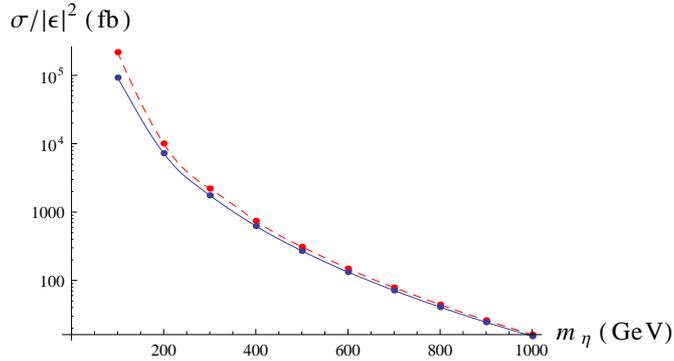}
\caption{Inclusive cross-sections normalised to the heavy-light neutrino mixing magnitude squared, $\eps^2$, for the charged current $pp \ra \eta l^\pm$ (blue) and neutral current $pp \ra \eta\nu$ (red) heavy neutrino production processes at the LHC for $s = 14$~TeV.}
\label{fig:pp2Nl}
\end{figure}

The production cross-section at the LHC is obtained after a convolution with the parton distribution functions. In Fig.~\ref{fig:pp2Nl}, we show the inclusive production cross-section obtained using \textsc{MadGraph 5}~\cite{Madgraph} for the LHC at 14~TeV CM energy. The production cross-section is normalised to the heavy-light neutrino mixings magnitude squared, $|\eps|^2$.
The on-shell heavy Majorana neutrino, $\eta$, subsequently decays into $Z\nu$, $W^\pm l^\mp$, $h_{SM}\nu$ with branching fractions roughly $20\%$, $40\%$, $40\%$ respectively. It is very unlikely to produce a heavy Majorana neutrino at LHC in the Type-I seesaw scenario, given that it would be very heavy 
($m_\eta \gg 1$~TeV), and the mixings involved are very small. For the inverse seesaw scenario however, since the heavy Majorana neutrino can be relatively light at a few hundred GeV and the mixing relatively large, it is possible to probe directly this scenario at LHC14, which is expected to have a luminosity of $\mathcal{O}(100)\,\mathrm{fb}^{-1}$. In particular, we see from Fig.~\ref{fig:pp2Nl} that for $m_\eta \sim 100$~GeV, one could have $\sigma \sim 1$~fb if
$|\eps|^2 \sim 10^{-5}$, which is easily obtainable in inverse seesaw. We leave a detailed study of the collider signals involved to future works.

\section{\label{sec:summ} Summary}
We have studied in this letter the implications of identifying the Majoron as DR. Assuming that it goes out of equilibrium at the muon annihilation temperature, it can account for the fractional value of the effective neutrino species, $N_\mathrm{eff}$, measured recently by Planck.
The consequence of this for the extended scalar sector associated with the Majoron is the presence of a very light scalar boson with mass $\lesssim 1.05$~GeV that mixes with the SM Higgs. Furthermore, the scale of the extended scalar sector, $v_S$, cannot be too high if excessive fine tuning of the parameters in the scalar potential is to be avoided.

The scalar sector scale, $v_S$, also sets the scale for the neutrino sector. A relatively low $v_S$ would however cause tension with the canonical Type-I seesaw scenario, which typically require a heavy scale above $10^{12}$ GeV. On the other hand, such tension would not arise in the inverse seesaw scenario. There, one can taking $v_S$ to be as low as a TeV without fine tuning either the Yukawa couplings or the scalar parameters, although consistency with the current LHC data
then requires the mixing between the scalar singlet and the Higgs doublet to be very small. This then implies that the corrections to the SM Higgs couplings will be not measurable at the LHC.

Low energy physics can provide further constraints on the light scalar mass. By demanding that the light scalar account for discrepancy between the experimental and theory value of the muon magnetic moment while consistent with the LHC data, the light scalar mass is pushed down to below 0.02~GeV. Although not able to completely solve the proton radius puzzle on its own, the light scalar can contribute a significant amount towards the 310~$\mu$eV of the muonic hydrogen Lamb shift required if it is even light with mass at around 1~MeV.

Finally, we are hopeful that the inverse seesaw scenario may be directly probed at LHC14 given that the heavy neutrino can be relatively light and the mixing relatively large.

\section{Acknowledgement}
The work of WFC is supported by Taiwan NSC under Grant No.102-2112-M-007-014-MY3. In addition, WFC gratefully acknowledges
the hospitality of TRIUMF Theory Group where part of this work was done. JNN is partially supported by the Natural Science and
Engineering Council of Canada, and would like to thank Prof. X.-G. He for his kind hospitality at the National Center for Theoretical Sciences where this work is completed. JMSW is supported by the National Center for Theoretical Sciences.


\end{document}